\documentstyle[prd,aps,eqsecnum,epsf,array]{revtex}

\begin{document}

\setlength{\topmargin}{-2ex}

\draft

\title{Stability of Excited Atoms in Small Cavities}

\author{ G. Flores-Hidalgo$\;$\thanks{E-mail:
gflores@cbpf.br}, A.P.C. Malbouisson$\;$\thanks{E-mail:
adolfo@cbpf.br}$\;$ and Y.W. Milla$\;$\thanks{E-mail:
yonym@cbpf.br}$\;$}  
\vspace{0.1cm} 
\address {\it Centro Brasileiro de Pesquisas F{\'\i}sicas, 
Rua Dr. Xavier Sigaud 150, Urca, Rio de Janeiro CEP 22290-180-RJ, Brazil.} 
  
\maketitle
 
\begin{abstract}
We consider a system consisting of an atom in the approximation of a harmonic 
oscillator of frequency $\bar{{\bf \omega}}$, coupled to the scalar potential inside a 
spherical reflecting cavity of  radius $R$.  We use dressed states introduced 
in a previous publication [Andion, Malbouisson, and Mattos Neto, J. Phys. A {\bf 34},
3735 (2001)], which allow a non-perturbative unified 
description of the atom radiation process, in both cases, of a finite or an 
arbitrarily large cavity. We perform a study of the energy distribution in a small 
cavity, with the initial condition that the atom is in the first excited state  
and we conclude for the quasi-stability of the excited atom. For instance, 
for a frequency $\bar{\omega}$ of the order $\bar{\omega}\sim 4.00\times 10^{14}/s$ 
(in the visible red), starting from the initial condition that the atom is in 
the first excited level, we find that for a cavity with diameter $2R\sim 1.0\times 
10^{-6}m$, the probability that the atom be at any time still in the first excited 
level, will be of the order of $97\%$. For  a typical microwave frequency 
$\bar{\omega}\sim 2,00\times 10^{10}/s$ we find stability in the first excited 
state also of the order of $97\%$ for a cavity radius $R\sim 1.4\times 10^{-2}m$.

\vspace{0.34cm}
\noindent
PACS Number(s):~03.65.Ca, 32.80.Pj 
 
\end{abstract}

\section{Introduction}
There are situations in the domain of Atomic Physics, Cavity Electrodynamics and
Quantum Optics, where perturbation methods are of little usefulness, for instance,  
in resonant effects associated to the coupling of atoms with strong radiofrequency 
fields \cite{bouquinCohen}. The theoretical understanding of these effects on  
perturbative grounds requires the calculation of very high-order terms in 
perturbation series, what makes the standard Feynman diagrams technique practically 
unreliable in those cases \cite{bouquinCohen}. The trials of treating 
non-perturbativelly such kind of systems, have lead to the idea of {\it dressed} 
atom, introduced in refs \cite{Polonsky} and \cite{Haroche}. Since then this concept 
has been  used  to investigate several situations involving the interaction of atoms 
and electromagnetic fields (\cite{Cohen1}, \cite{Cohen2}, \cite{Haroche1}). A way to 
circumvect the mathematical difficulties due to non-linear character of the problem, 
is to assume that under certain conditions the coupled atom-electromagnetic field  
system may be approximated by the system composed of an harmonic oscillator coupled 
{\it linearly} to the field trough some effective coupling constant $g$. This is the 
case in the context of the general $QED$ linear response theory, where the electric 
dipole interaction gives the leading contribution to the radiation process 
(\cite{McLachlan}, \cite{Wylie}, \cite{Jhe}, \cite{Passante}) and also in several 
branches of Quantum Optics (\cite{Jmario}, \cite{Davidovitch}, \cite{Fonseca}). 
  
We consider a system composed of an atom (approximated by a harmonic oscillator) 
coupled linearly to the scalar potential, the whole system being confined inside a 
reflecting sphere of radius $R$. We give a non-perturbative treatment to the field-atom
system introducing some {\it dressed} coordinates that allow to divide the coupled 
system into two parts, the {\it dressed} atom and the {\it dressed} field, what makes
unnecessary to work directly with the concepts of bare atom, bare field and interaction
between them. For instance, to describe the radiation process, having as initial 
condition that only the mechanical oscillator (the atom), $q_{0}$ be excited, the usual
procedure is to consider the interaction term in the Hamiltonian written in terms of 
$q_{0}$ and the field modes $q_{i}$ as a perturbation, which induces transitions among
the eigenstates of the free Hamiltonian. In this way it is possible to treat 
approximately the problem having as initial condition that only the bare mechanical 
oscillator (the atom) be excited. But as is well known this initial condition is 
physically not consistent in reason of the divergence of the bare oscillator frequency, 
due to the interaction with the field. The traditional way to circumvect this difficulty 
is by the renormalization procedure, introducing perturbativelly order by order 
corrections to the oscillator frequency. In this paper we adopt an alternative procedure
introduced in \cite{nelson}. We do not make explicit use of the concepts of interacting 
bare oscillator and field, described by the coordinates $q_{0}$ and $\{q_{i}\}$. We 
introduce {\it dressed} coordinates $q^{\prime}_{0}$ and $\{q^{\prime}_{i}\}$ for, 
respectively the {\it dressed} atom and the {\it dressed} field modes. In terms of 
these new coordinates a non-perturbative approach of the radiation process and of the 
distribution of energy inside the cavity is possible.

We start considering an atom approximated by a harmonic oscillator $q_{0}(t)$ of 
frequency $\omega_{0}$ (we will introduce below a {\it renormalized} frequency 
$\bar{\omega}$ which is physically meaningfull) coupled linearly to  the scalar 
potential $\phi$, the whole system being confined in a sphere of radius $R$ centered 
at the oscillator position. The equations of motion are,  
\begin{equation} 
\mbox{\"q}_{0}(t)+\omega_{0}^{2}q_{0}(t) = 2\pi \sqrt{gc}\int_{0}^{R}d^{3} 
{\bf r}\phi({\bf r},t)\delta ({\bf r})\;, 
\label{eq. mov5} 
\end{equation} 
\begin{equation} 
\frac{1}{c^{2}}\frac{\partial^{2}\phi}{\partial t^{2}}-\nabla^{2} \phi({\bf r 
},t)=2\pi\sqrt{gc}q_{0}(t)\delta({\bf r})\;.  
\label{mov1} 
\end{equation}
Using a basis of spherically symmetric Bessel functions defined in the domain $0<| 
{\bf r}|<R$, the equations above can be written as a set of equations coupling the atom
to the harmonic field modes, which can be derived from the Hamiltonian
\begin{equation} 
H=\frac{1}{2}\left[p_{0}^{2}+\omega_{0}^{2}q_{0}^{2}+ 
\sum_{k=1}^{N}(p_{k}^{2}+\omega_{k}^{2}q_{k}^{2}-2\eta\omega_{k}q_{0}q_{k}) 
\right]\;.
\label{Hamiltoniana} 
\end{equation}

\section{The eigenfrequencies spectrum}

We consider for a moment as in \cite{nelson}, the problem of a harmonic oscillator 
$q_{0}$ coupled to $N$ other oscillators. In the limit $N\rightarrow \infty$ we recover
our original situation of the coupling oscillator-field after redefinition of divergent
quantities, in a manner analogous as renormalization is done in field theories. In the 
above equations, $g$ is a coupling constant (with dimension of frequency), $\eta=
\sqrt{2g\Delta\omega}$ and $\Delta\omega=\pi c/R$ is the interval between two 
neighbouring field frequencies, $\omega_{i+1}-\omega_{i}=\Delta\omega=\pi c/R $ and 
$q_{i}$ stands for the harmonic modes of the field. The Hamiltonian (\ref{Hamiltoniana})
can be turned to principal axis by means of a point transformation,   
$q_{\mu}=t_{\mu}^{r}Q_{r}\,,\,\,\,p_{\mu}=t_{\mu}^{r}P_{r}$,  
performed by an orthonormal matrix $T=(t_{\mu}^{r})$, \, $\mu=(0,k)$, \, $ 
k=1,2,...\, N$, $r=0,...N$. The subscript $0$ and $k$ refer respectively to the atom and 
the harmonic modes  of the field and $r$ refers to the normal modes. The transformed 
Hamiltonian in principal axis reads,   
$H=\frac{1}{2}\sum_{r=0}^{N}(P_{r}^{2}+\Omega_{r}^{2}Q_{r}^{2})$,  
where the $\Omega_{r}$'s are the normal frequencies corresponding to the possible 
collective oscillation modes of the coupled system. The eigenfrequencies $\Omega_{r}$ 
satisfy the equation \cite{nelson},
\begin{equation} 
\omega_{0}^{2}-N\eta^{2}-\Omega^{2}=\eta^{2}\sum_{k=1}^{N}\frac{\Omega^{2}}{ 
\omega_{k}^{2} -\Omega^{2}}\;.  
\label{Nelson1} 
\end{equation} 
The $N+1$ solutions $\Omega_{r}$ of Eq.(\ref{Nelson1}), correspond to the $N+1$ normal 
collective oscillation modes.

It can be shown \cite{nelson} that if $\omega_{0}^{2}>N\eta^{2}$, Eq.(\ref{Nelson1}) 
yields only positive solutions for $\Omega^{2}$  (all collective modes are harmonic),
while if $\omega_{0}^{2}<N\eta^{2}$, Eq.(\ref{Nelson1}) has a negative solution  
$\Omega_{-}^{2}$. This means that there is a damped oscillation mode that does not 
allows stationary configurations. We will not care about this last situation. 
Nevertheless it should be remarked that in a different context, it is precisely this 
negative squared frequency solution (runaway solution) that is related to the 
existence of a bound state in the Lee-Friedrechs model. This solution is considered 
in ref. \cite{Likhoded} in the framework of a model to describe qualitatively the 
existence of bound states in particle physics. Thus we take $\omega_{0}^{2}>N\eta^{2}$
and define the {\it renormalized} oscillator frequency $\bar{\omega}$,   
$\bar{\omega}=\sqrt{\omega_{0}^{2}-N\eta^{2}}$. In the limit $N\rightarrow \infty$ the
meaning of the frequency renormalization becomes clear. It is exactly the analogous of 
a mass renormalization in field theory, the infinite $\omega_{0}$ being chosen in such 
a way as to make the renormalized frequency $\bar{\omega}$ finite and equal to the 
observed oscillator frequency. In terms of the renormalized frequency Eq.(\ref{Nelson1})
can be writen, after some manipulations, in the form \cite{nelson}, 
\begin{equation} 
\mathrm{cot}(\frac{R\Omega}{c})=\frac{\Omega}{\pi g}+\frac{c}{R\Omega}
(1-\frac{\bar{\omega}^{2}R}{\pi gc})\;.  
\label{eigenfrequencies} 
\end{equation}
The solutions of Eq.(\ref{eigenfrequencies})  
 with respect 
to $\Omega$ give the spectrum of eigenfrequencies $\Omega_{r}$ corresponding to the 
collective normal modes. The transformation matrix elements turning the oscillator-field 
system to principal axis is obtained taking the limit $N\rightarrow \infty$, after some 
rather long but straightforward manipulations in \cite{nelson}. They read,  
\begin{eqnarray} 
t_{0}^{r}&=&\frac{\Omega_{r}}{\sqrt{\frac{R}{2\pi gc}(\Omega_{r}^{2}-\bar{ 
\omega}^{2})^{2}+\frac{1}{2}(3\Omega_{r}^{2}-\bar{\omega}^{2})+\frac{\pi gR}{ 
2c}\Omega_{r}^{2}}}\;,\nonumber\\   
t_{k}^{r}&=&\frac{\eta\omega_{k}}{\omega_{k}^{2}-\Omega_{r}^{2}}t_{0}^{r}\;. 
\label{t0r2} 
\end{eqnarray} 
 
We define below some coordinates $q^{\prime}_{0}$, $q^{\prime}_{i}$ associated  to the 
{\it dressed} atom and the {\it dressed} field. These coordinates will reveal themselves
to be appropriate to give an appealing non-perturbative description of the atom-field 
system.

\section{Dressed states}

We start from the eigenstates of our system, represented by the normalized eigenfunctions,  
\begin{equation} 
\phi_{n_{0}n_{1}n_{2}...}(Q,t)=\prod_{s}\left[\sqrt{\frac{2^{n_s}}{n_s!}}
H_{n_{s}}(\sqrt{\frac{ \Omega_{s}}{\hbar}}Q_{s})\right]
\Gamma_{0}e^{-i\sum_{s}n_{s}\Omega_{s}t}\;, 
\label{autofuncoes} 
\end{equation} 
where $H_{n_{s}}$ stands for the $n_{s}$-th Hermite polynomial and $\Gamma_{0}$ is the 
normalized vacuum eigenfunction. Let us introduce {\it dressed} coordinates 
$q^{\prime}_{0}$ and $\{q^{\prime}_{i}\}$ for, respectively the {\it dressed} atom and  
the {\it dressed} field, defined by \cite{nelson},  
\begin{equation} 
\sqrt{\frac{\bar{\omega}_{\mu}}{\hbar}}q^{\prime}_{\mu}=\sum_{r}t_{\mu}^{r} 
\sqrt{\frac{\Omega_{r}}{\hbar}}Q_{r}\;,  
\label{qvestidas1} 
\end{equation} 
valid for arbitrary $R$ and where $\bar{\omega}_{\mu}=\{\bar{\omega}, \;\omega_{i}\}$. 
In terms of the bare coordinates the dressed coordinates are expressed as,  
\begin{equation} 
q^{\prime}_{\mu}=\sum_{\nu}\alpha_{\mu \nu}q_{\nu}\;;  \;\;\;\;\;\alpha_{\mu \nu}=
\frac{1}{\sqrt{\bar{\omega}_{\mu}}}\sum_{r}t_{\mu}^{r}t_{ \nu}^{r}\sqrt{\Omega_{r}}\;. 
\label{qvestidas3} 
\end{equation} 
Let us define for a fixed instant the complete orthonormal set of functions \cite{nelson},  
\begin{equation} 
\psi_{\kappa_{0} \kappa_{1}...}(q^{\prime})=\prod_{\mu}\left[\sqrt{\frac{2^{\kappa_{\mu}}}
{\kappa_{\mu}!}} H_{\kappa_{\mu}} (\sqrt{\frac{\bar{\omega}_{\mu}}{\hbar}} 
q^{\prime}_{\mu})\right]\Gamma_{0}\;,  
\label{ortovestidas1} 
\end{equation} 
where $q^{\prime}_{\mu}=q^{\prime}_{0},\, q^{\prime}_{i}$, $\bar{\omega}_{\mu}=
\{\bar{\omega},\, \omega_{i}\}$. Note that the ground state $\Gamma_{0}$ in the above 
equation is the same as in Eq.(\ref{autofuncoes}). The invariance of the ground state is 
due to our definition of {\it dressed} coordinates given by Eq.(\ref{qvestidas1}). Each 
function $\psi_{\kappa_{0} \kappa_{1}...}(q^{\prime})$ describes a state in which the 
{\it dressed} oscillator $q'_{\mu}$ is in its $\kappa_{\mu}-th$ excited state. Using 
Eq.(\ref{qvestidas1}) the functions (\ref{ortovestidas1}) can be expressed in terms of 
the normal coordinates $Q_{r}$. But since (\ref{autofuncoes}) is a complete set of 
orthonormal functions, the functions (\ref{ortovestidas1}) may be written as linear 
combinations of the eigenfunctions of the coupled system (we take $t=0$ for the moment),  
\begin{equation} 
\psi_{\kappa_{0} 
\kappa_{1}...}(q^{\prime})=\sum_{n_{0}n_{1}...}T_{\kappa_{0} \kappa_{1}...} 
^{n_{0}n_{1}...}(0)\phi_{n_{0}n_{1}n_{2}...}(Q,0)\;,  
\label{ortovestidas2} 
\end{equation} 
where the coefficients are given by,  
\begin{equation} 
T_{\kappa_{0} \kappa_{1}...}^{n_{0}n_{1}...}(0)=\int dQ\, \psi_{\kappa_{0} 
\kappa_{1}...}\phi_{n_{0}n_{1}n_{2}...}\;, 
 \label{ortovestidas3} 
\end{equation} 
the integral extending over the whole $Q$-space.

We consider the particular configuration $\psi$ in which only one dressed oscillator
$q^{\prime}_{\mu}$ is in its $N$-th excited state, all other being in the ground state, 
\begin{equation} 
\psi_{0...N(\mu)0...}(q^{\prime})=(2^{-N}N!)^{-\frac{1}{2}} H_{N}
(\sqrt{\frac{\bar{\omega}_{\mu} }{\hbar}}q^{\prime}_{\mu})\Gamma_{0}\;.  
\label{ortovestidas4} 
\end{equation} 
The coefficients (\ref{ortovestidas3}) have been calculated in ref.\cite{nelson}. We 
get,  
\begin{equation} 
T_{0...N(\mu)0...}^{n_{0}n_{1}...}=(\frac{N!}{n_{0}!n_{1}!...})^{\frac{1}{2}%
}(t_{\mu}^{0})^{n_{0}}(t_{\mu}^{1})^{n_{1}}...\;,  
\label{coeffN} 
\end{equation} 
where the subscripts $\mu=0,\; i$ refer respectively to the dressed atom and the harmonic 
modes of the field and the quantum numbers satisfy the constraint $n_{0}+n_{1}+...=N$.
In the following we focus our attention on the behaviour of the system with the initial 
condition that only one dressed  oscillator $q^{\prime}_{\mu}$ (the dressed atom or one 
of the modes of the dressed field) be in the $N$-th excited state. We will study in
detail the particular case $N=1$, which will be enough to have a clear understanding of 
our approach. Let us call $\Gamma_{1}^{\mu}$ the configuration in which only the dressed 
oscillator $q^{\prime}_{\mu}$ is in the first excited level. We have from 
Eqs.(\ref{ortovestidas4}), (\ref{ortovestidas2}) (\ref{coeffN}) and (\ref{qvestidas1}) 
the following expression for the time evolution of the first-level excited dressed 
oscillator $q^{\prime}_{\mu}$,  
\begin{equation} 
\Gamma_{1}^{\mu}(t)=\sum_{\nu}f^{\mu \nu}(t)\Gamma_{1}^{\nu}(0)\;, 
\label{ortovestidas5} 
\end{equation} 
where the coefficients $f^{\mu \nu}(t)$ are given by  
\begin{equation} 
f^{\mu \nu}(t)=\sum_{s}t_{\mu}^{s}t_{\nu}^{s}e^{-i\Omega_{s}t}\;. 
\label{fmunu} 
\end{equation} 
From Eq.(\ref{ortovestidas5}) we see that the initially excited dressed oscillator 
naturally distributes its energy among itself and all other dressed oscillators as time 
goes on, with probability amplitudes given by Eq.(\ref{fmunu}). If the dressed oscillator 
$q'_{0}$ (the atom) is in its first excited state at $t=0$, its decay rate may evaluated 
from the time evolution equation,  
\begin{equation} 
\Gamma_{1}^{0}(t)=\sum_{\nu}f^{0 \nu}(t)\Gamma_{1}^{\nu}(0)\;. 
\label{ortovestidas6} 
\end{equation} 
In Eq.(\ref{ortovestidas6}) the coefficients $f^{0 \nu}(t)$ have a simple interpretation: 
$f^{00}(t)$ and $f^{0i}(t)$  are respectively the probability amplitudes that at time 
$t$ the dressed atom still be excited or have radiated a photon of frequency $\omega_{i}$.
We see that this formalism allows a quite natural description of the radiation process as 
a simple exact time evolution of the system. We consider in the following the time 
evolution of the excited atom, in the cases of a very large and a very small cavity.

{\it A very large cavity}:

In the case of a very large cavity our method generalizes what can be obtained from 
perturbation theory. The  probability that the atom be still excited at time $t$ 
can be obtained in continuous language from the amplitude given by Eq. (\ref{fmunu}),  
\begin{equation} 
f^{00}(t)=\int_{0}^{\infty}\frac{2g\Omega^{2}e^{-i\Omega t}\, d\Omega} 
{(\Omega^{2}-\bar{\omega}^{2})^{2}+\pi^{2}g^{2}\Omega^{2}}\;.
\label{f00} 
\end{equation} 
For large $t$ ($t>> \frac{1}{\bar{\omega}}$), but for in principle arbitrary coupling 
$g$, we obtain for the probability of finding the atom still excited at time $t$, the 
result \cite{nelson},  
\begin{eqnarray} 
|f^{00}(t)|^{2}=e^{-\pi gt}(1+\frac{\pi^{2}g^{2}}{4\tilde{\bar{\omega}}^{2}})-
e^{-\pi gt/2} \frac{8 g}{ \bar{\omega}^{4}t^{3}}(\sin\tilde{\bar{\omega} 
}t+\frac{\pi g}{2\tilde{\bar{\omega}}} \cos\tilde{\bar{\omega}}t)+
\frac{16g^{2}}{\bar{\omega}^{8}t^{6}}\;,   
\label{|f00|2} 
\end{eqnarray} 
where $\tilde{\bar{\omega}}=\sqrt{\bar{\omega}^{2}-\frac{\pi^{2}g^{2}}{4}}$. In the 
above expression the approximation $t>>\frac{1}{\bar{\omega}}$ plays a role only in the
two last terms, due to difficulties to evaluate exactly the integral in Eq. (\ref{f00}) 
along the imaginary axis using Cauchy's theorem. The first term comes from the residue 
at $\Omega=\tilde{\bar{\omega}}+i\frac{\pi g}{2}$ and would be the same if we have done
an exact calculation. If we consider in eq. (\ref{|f00|2}) $g<<\bar{\omega}$, which 
corresponds in electromagnetic theory to the fact that the fine structure constant 
$\alpha$ is small compared to unity (for explicit calculations we take below 
$g/\bar{\omega}=\alpha)$, we obtain the well known perturbative exponential decay law.

\section{The radiation process in a small cavity}

Let us now consider the atom placed at the center of a very small cavity, $i.e.$ that
satisfies the condition that its radius be much smaller than the coherence lenght, 
$R<<c/g$. To obtain the eigenfrequencies spectrum, we remark that from a numerical 
analysis of Eq.(\ref{eigenfrequencies}) it can be seen that in the case of a small 
cavity radius $R$, its  solutions are near the frequency values corresponding to the 
asymptots of the curve $\mathrm{cot}(\frac{R\Omega}{c})$, which correspond to the 
field modes $\omega_{i}=i\pi c/R$. The smallest solution departs more from the first
asymptot than the other larger solutions depart from their respective nearest asymptot.
As we take larger and larger solutions, they are nearer and nearer to the values 
corresponding to the asymptots. For instance, for a cavity radius $R$ of the order of
$10^{-2}m$ and $\bar{\omega}\sim 10^{10}/s$, only the lowest eigenfrequency 
$\Omega_{0}$ is signicantly different from the field frequency corresponding to the
first asymptot, all the other eigenfrequencies $\Omega_{k}, \;k=1,2,...$ being very 
close to the field modes $k\pi c/R$. For higher values of $\bar{\omega}$ (and lower 
values of $R$) the differences between the eigenfrequencies and the field modes 
frequencies are still smaller.

Thus to solve Eq.(\ref{eigenfrequencies}) for the larger eigenfrequencies we expand 
the function $\mathrm{cot}(\frac{R\Omega}{c})$ around the values corresponding to the
asymptots. We write,
\begin{equation}
\Omega_k=\frac{\pi c}{R}(k+\epsilon_k)\;,~~~k=1,2,..
\label{others}
\end{equation} 
with $0<\epsilon_{k}<1$, satisfying the equation,
\begin{equation}
\mathrm{cot}(\pi \epsilon_k)=\frac{c}{gR}(k+\epsilon_k)+
\frac{1}{(k+\epsilon_k)}(1-\frac{\bar{\omega}^{2}R}{\pi gc})\;.  
\label{eigen2} 
\end{equation}
But since for a small cavity every $\epsilon_k$ is  much smaller than $1$, 
Eq.({\ref{eigen2}) may be linearized in  $\epsilon_k$, giving, 
\begin{equation}
\epsilon_k=\frac{\pi g c R k}{\pi^{2} c^{2} k^{2}-\bar{\omega}^{2}R^{2}}\;.
\label{linear}
\end{equation}
Eqs.(\ref{others}) and (\ref{linear}) give approximate solutions to the 
eigenfrequencies $\Omega_{k},\;\;k=1,2...$.

To solve Eq.(\ref{eigenfrequencies}) with respect to the lowest eigenfrequency 
$\Omega_{0}$, let us assume that it satisfies the condition $\Omega_{0}R/c<<1$ (we 
will see below that this condition is compatible with the condition of a small cavity
as defined above). 
Inserting the condition $\Omega_{0}R/c<<1$ in Eq.(\ref{eigenfrequencies})and keeping 
up to quadratic terms in $\Omega$ the solution for the lowest eigenfrequency  
$\Omega_{0}$ can be writen,
\begin{equation}
\Omega_0=\frac{\bar{\omega}}{\sqrt{1+\frac{\pi gR}{c}}}\;.
\label{firsts}
\end{equation}
Consistency between Eq.(\ref{firsts}) and the condition  $\Omega_{0}R/c<<1$ gives 
a condition on the cavity radius,
\begin{equation}
R\ll 
\frac{c}{g}\frac{\pi}{2}\left(\frac{g}{\bar{\omega}}\right)^2
\left(1+\sqrt{ 1+\frac{4}{\pi^2}\left(\frac{\bar{\omega}}{g}\right)^2}~\right)\;.
\label{rsmall}
\end{equation}
Let us define the coupling constant $g$ to be such that $g=\bar{\omega}\alpha$, where 
$\alpha$ is the fine structure constant, $\alpha=1/137$. Then the factor multiplying 
$c/g$ in the above equation is $\sim 0.07$ and the condition $R\ll c/g$ is replaced 
by a more restrictive assumption $R\ll 0.07(c/g)$. For a typical infrared frequency, 
for instance $\bar{\omega}\sim 2,0\times 10^{11}/s$, our calculations are valid for a
radius $R\ll 10^{-3}m$.

From Eq.(\ref{t0r2}) and using the above expressions for the eigenfrequencies in a 
small cavity, we obtain the matrix elements,
\begin{equation}
(t_0^0)^2\approx 1-\frac{\pi g R}{c};\;\;(t_0^k)^2\approx \frac{2 g R}{\pi c k^2}\;.
\label{too}
\end{equation}
To obtain the above equations we have neglected the corrective term $\epsilon_{k}$,
from the expressions for the eigenfrequencies $\Omega_{k}$.   
Nevertheless, corrections in $\epsilon_{k}$ should be included in the expressions for
the matrix elements $t_{k}^{k}$, in order to avoid spurious singularities due to our 
approximation.

Let us consider the situation where the dressed atom is initially in its first 
excited level. Then from Eq.(\ref{fmunu}) we obtain the probability that it will 
still be excited after a ellapsed time $t$, 
\begin{equation}
|f^{00}(t)|^{2}=(t_{0}^{0})^{4}+2\sum_{k=1}^{\infty}
(t_{0}^{0})^{2}(t_{0}^{k})^{2}\cos(\Omega_{k}-\Omega_{0})t+
\sum_{k,l=1}^{\infty}(t_{0}^{k})^{2}
(t_{0}^{l})^{2}\cos(\Omega_{k}-\Omega_{l})t\;.
\label{|f00R|2}
\end{equation}
Using Eqs.(\ref{too}) in Eq.({\ref{|f00R|2}), we obtain 
\begin{equation}
|f^{00}(t)|^{2}=1-\pi \delta+4(\frac{\delta}{\pi}-\delta^{2})
\sum_{k=1}^{\infty}\frac{1}{k^{2}}\cos(\Omega_{k}-\Omega_{0})t
\pi^{2}\delta^{2}+\frac{4}{\pi^{2}}\delta^{2}
\sum_{k,l=1}^{\infty}\frac{1}{k^{2}l^{2}}\cos (\Omega_{k}-\Omega_{l})t\;,
\label{f002}
\end{equation} 
where we have introduced the adimensional parameter $\delta=Rg/c\;\ll 1$, 
corresponding to a small cavity and we remember that the eigenfrequencies are given
by Eqs.(\ref{others}) and (\ref{linear}). As time goes on, the probability that the
atom be excited  attains periodically a minimum value which has a lower bound given
by,
\begin{equation}
\mathrm{Min}(|f^{00}(t)|^{2})=1-\frac{5\pi}{3}\delta+\frac{14\pi^{2}}{9}\delta^{2}\;.
\label{min}
\end{equation}
For a frequency $\bar{\omega}$ of the order $\bar{\omega}\sim 4.00\times 10^{14}/s$ 
(in the red visible),  
which corresponds to $\delta\sim 0.005$ and $2R\sim 1.0\times 10^{-6}m$, we see from
Eq.(\ref{min}) that the probability that the atom be at any time excited will never 
fall below a value $\sim 0.97$, or a decay probability that is never higher that a 
value $\sim 0.03$. In other words, atoms having such emission frequency, placed in a 
such a small cavity in the first excited level, will be stable in the excited state 
to the order of $97\%$. It is interesting to compare this result with experimental 
observations in \cite{Haroche3}, \cite{Hulet}, where stability is found for atoms 
emiting in the visible range placed between two parallel mirrors a distance 
$L=1.1\times 10^{-6}m$ apart from one another. For lower frequencies the size of the
cavity ensuring quasi-stability of the same order as above, for the excited atom may 
be considerably larger. For instance, for $\bar{\omega}$ in a typical microwave 
value, $\bar{\omega}\sim 2,00\times 10^{10}/s$ and taking also $\delta \sim 0.005$,  
the probability that the atom remain in the first excited level at any time will be 
larger than a value of the order of $97\%$, for a cavity radius 
$R\sim 1.0\times 10^{-2}m$. The probability that the atom remain excited as time goes 
on oscillates with time between a maximum and a minimum values and never departs 
significantly from the situation of stability of the atom in the excited state.
Indeed for an emission frequency $\bar{\omega}\sim 4.00\times 10^{14}/s$ (in the red
visible) considered above and $R\sim 1.0\times 10^{-6}m$, the period of oscillation 
between the minimum and maximum values of the probability that the atom be excited,  
is $T\sim \frac{1}{12}\times 10^{-14}s$, while for 
$\bar{\omega}\sim 2,00\times 10^{10}/s$, and $R\sim 1.4\times 10^{-2}m$, the period 
is $T\sim \frac{1.4}{6}\times 10^{-10}s$.

\section{Concluding remarks}
 
We have used in this paper a formalism that allows an unified approach to the 
radiation process by an atom, in rather different situations, as the atom confined in
a very small cavity or in free space. The behaviour of atoms confined in small cavities 
is completelly different from the behaviour of an atom in free space or in a large 
cavity. In the first case the emission process is very sensitive to the presence of 
boundaries, a fact that has been pointed out since a long time ago in the literature
(\cite{Morawitz}, \cite{Milonni}, \cite{Kleppner}). Our {\it dressed} states approach 
gives an unified description for the dressing of the atom by the field modes and the 
emission process in a cavity of arbitrary size, which includes microcavities and very 
large cavities (free space emission). We recover here with our formalism the 
experimental observation that excited states of atoms in sufficiently small cavities 
are stable. We are able to give formulas for the probability of an atom to remain 
excited for an infinitely long time, provided it is placed in a cavity of appropriate
size. For an emission frequency in the visible red, the size of such cavity is in good 
agreement with experimental observations (\cite{Haroche3}, \cite{Hulet}). Also, our 
approach gives results in good agreement with previous theoretical results for the 
emission in free space, generalizing the well known exponential decay law. Moreover 
the detailed behaviours which we obtain with our formalism are very different in the 
two situations: The atom in a very large cavity has a probability decay rate weekly 
oscillating and monotonically varying with time (the probability that the atom be 
excited decreases almost exponentially with increasing time). In the case of an 
excited atom placed in the center of a very small cavity, the probability that it 
remains excited as time goes on oscillates very rapidly with time and never departs 
significantly from the situation of stability of the atom in the excited state.

\section{Acknowledgements}

This work received financial support from CNPq (Brazilian National Research Council)
and FAPERJ (Foundation for the Support of Research in the State of Rio de Janeiro 
(Brazil)). One of us (A.P.C.M.) is grateful to C. de Calan (Ecole Polytechnique, 
Paris), F.S. Nogueira and H. Kleinert (Freie Universit\"at, Berlin) for interesting 
discussions.

\end{document}